\documentclass[11pt, oneside]{article}
\usepackage[a4paper, total={6.2in, 9in}]{geometry}
\pdfoutput=1
\usepackage{graphicx}
\usepackage{amssymb}
\usepackage{xcolor}
\usepackage[T1]{fontenc}
\usepackage{lmodern}
\usepackage{makeidx}
\usepackage[normalem]{ulem}

\PassOptionsToPackage{hyphens}{url}\usepackage[hidelinks]{hyperref}
\usepackage{amsmath}

\usepackage{fancyhdr}                           % Needed for footers on each page
\fancypagestyle{firstpage} % first page footer only
{
   \fancyhf{}
   \fancyfoot[C]{\vspace{-12mm}\thepage}

   \fancyfoot[L]{\vspace{-5mm}\noindent\rule{6.2in}{0.4pt} 
     \scriptsize{\textcopyright \hspace{0.5mm} Barclays Bank PLC 2016 \\
       This work is licensed under a \href{https://creativecommons.org/licenses/by/4.0/}{Creative Commons Attribution 4.0 International License (CC
       BY).}  \\ Provided you adhere to the CC BY license, including as to attribution, you are
       free to copy and redistribute this work in any medium or format and remix, transform,
       and build upon the work for any purpose, even commercially.                  \\
       BARCLAYS is a registered trade mark of Barclays Bank PLC, all rights are reserved. \\
       \vspace{2mm}
       Electronic copy is available at \url{https://arxiv.org/abs/1612.04496}
   }
 }
}

\title{\vspace{-1cm}Smart Contract Templates: \\essential requirements and design options}

\usepackage{anyfontsize}
\author{%
  \hspace{-1.5em}
  \begin{tabular}{c} {\fontsize{9.75}{1cm}\selectfont Christopher D. Clack} \\ {\fontsize{9.75}{1cm}\selectfont Centre for Blockchain
      Technologies} \\ {\fontsize{9.75}{1cm}\selectfont Department of Computer Science} \\ {\fontsize{9.75}{1cm}\selectfont University College
      London} \end{tabular} \hspace{-2.5em} \and
  \begin{tabular}{c} {\fontsize{9.75}{1cm}\selectfont Vikram A. Bakshi} \\ {\fontsize{9.75}{1cm}\selectfont Investment Bank CTO Office} \\
    {\fontsize{9.75}{1cm}\selectfont Barclays} \\ \hskip 1em \end{tabular} \hspace{-2em} \and
  \begin{tabular}{c} {\fontsize{9.75}{1cm}\selectfont Lee Braine} \\ {\fontsize{9.75}{1cm}\selectfont Investment Bank CTO Office} \\
    {\fontsize{9.75}{1cm}\selectfont Barclays}  \\ \hskip 1em \end{tabular} 
}

\date{December 15, 2016} 

\begin{document}
\maketitle
\thispagestyle{firstpage} % Needed to get footer on first page
\vspace{-1cm}
\begin{abstract}

  Smart Contract Templates support legally-enforceable smart contracts, using operational
  parameters to connect legal agreements to standardised code. In this paper, we explore the
  design landscape of potential formats for storage and transmission of smart legal
  agreements. We identify essential requirements and describe a number of key design options,
  from which we envisage future development of standardised formats for defining and
  manipulating smart legal agreements. This provides a preliminary step towards supporting
  industry adoption of legally-enforceable smart contracts.
\end{abstract}

\vspace{-5mm}

\section{Introduction}
\label{sec:introduction-and-scope}

The aim of Smart Contract Templates \cite{smartcontracttemplates,SCT2016} is to support the
management of the complete lifecycle of smart legal agreements. This includes the creation of
legal document templates by standards bodies and the subsequent use of those templates in the
negotiation and agreement of contracts by counterparties. They also facilitate automated
execution of the contract via smart contract code \cite{stark2016}, and provide a direct link
within the instantiated smart contract as an identifier for reference and recovery of the
signed legal agreement. The smart legal contracts could potentially be executed on distributed
ledgers (such as Axoni Core \cite{axoni}, \linebreak Corda \cite{corda}, Digital Asset Platform
\cite{digitalasset}, Ethereum \cite{ethereum}, Hyperledger Fabric \cite{hyperledger}, etc.).

In a previous paper \cite{SCT2016}, we discussed the foundations, design landscape and research
directions for Smart Contract Templates. We begin this paper by stating what we believe are the
essential requirements for smart legal agreements. We then provide an abstract ``core''
specification and proceed to explore the design landscape for the storage and transmission of
smart legal agreements. Our aim is to support the financial services industry (including trade
associations such as the International Swaps and Derivatives Association (ISDA) and FIA) in:
(i) exploring how legal prose can be connected with parameters and code, and (ii) reviewing
existing data standards to take account of the features of smart legal agreements.

We do not aim to address topics relating to the execution of smart contract code,
the semantics of legal prose, or languages for expressing business logic.

In a similar manner to our previous paper \cite{SCT2016}, we aim to discuss these topics using
reasonably straightforward language, so that it is accessible not only to financial
institutions but also to lawyers, standards bodies, regulators, and policy makers.  We hope
that the issues and views raised in this paper will stimulate debate and we look forward to
receiving feedback.

\vspace{1.5mm}
\noindent \textbf{Acknowledgements:} We would like to thank Clive Ansell (ISDA), Ian Grigg
(R3), Darren Jones (Barclays) and Simon Puleston Jones (FIA) for their helpful feedback.

\section{Essential requirements}
\label{sec:essential-requirements}
Smart Contract Templates are based on the framework of Grigg's Ricardian Contract triple of
``prose, parameters and code'' \cite{grigg2004ricardian, grigg2015sumofchains}. In this
framework, key operational parameters (hereafter called ``execution parameters'') are extracted
from the legal prose and passed to the smart contract code that provides automated 
execution.
  
The parameters are a succinct way to direct the code; additionally, one of those parameters may
be an identifier for the reference and recovery of the smart legal agreement. The aim is to
provide a legally-enforceable foundation for smart contracts (explained in more detail in
\cite{SCT2016}).

The above description leads to the
following {\em essential requirements} for smart legal agreements:

{\em
  \begin{enumerate}
    \item Methods to create and edit smart legal agreements, including legal prose and
      parameters.
    \item Standard formats for storage, retrieval and transmission of smart legal agreements.
    \item Protocols for legally executing smart legal agreements (with or without signatures).
    \item Methods to bind a smart legal agreement and its corresponding smart contract code to
      create a legally-enforceable smart contract.
    \item Methods to make smart legal agreements available in forms acceptable according to
      laws and regulations in the appropriate jurisdiction.
  \end{enumerate}
}

\noindent
The above essential requirements include four key items that merit further discussion:

\begin{enumerate}
\item
{\em Editing.} Lawyers are likely to favour a graphical What-You-See-Is-What-You-Get
(WYSIWYG\footnote{See https://en.wikipedia.org/wiki/WYSIWYG}) editor.  This may be an existing
ubiquitous editor (such as Microsoft Word), an editor enhanced with add-ins (such as Thomson
Reuters Contract Express Author \cite{contractexpress} or HotDocs \cite{hotdocs}), or a custom
editor (such as Smart Communications SmartDX \cite{smartdx} or ClauseMatch \cite{clausematch}).
Alternatives include text editors (which may or may not include syntax highlighting) and
graphical What-You-See-Is-What-You-Mean (WYSIWYM\footnote{See
  https://en.wikipedia.org/wiki/WYSIWYM}) editors. The editor must support contract metadata,
including parameters.

\item
{\em Transmission.} To facilitate the transmission of smart legal agreements between multiple
counterparties (e.g. during negotiation) and between a range of different applications
(e.g. agreement editors and analytical tools), there should be agreed standard formats for
transmission.

There are many possible ``concrete'' formats that could be used to transmit smart legal
agreements. These include formats based on Extensible Markup Language \linebreak (XML)
\cite{xml} (such as Office Open XML Document \cite{openxml} and Open Document Format for Office
Applications (ODF) \cite{opendocument}), JSON\footnote{JSON is used by, for example, Common
  Form \cite{commonform}} \cite{json}, markdown\footnote{See
  \url{https://en.wikipedia.org/wiki/Markdown}.  Markdown is used by, for example, CommonAccord
  \cite{commonaccord}.}, etc.  If a standard format for transmission is not utilised, then it
would be necessary, for example, to translate between formats during import and export and
semantic consistency may not be assured.

It may be necessary to have different concrete implementations for different product
categories, for example derivatives versus syndicated loans.  We propose there would be benefit
in a formal ``abstract'' specification for a serialised format. Such a specification could
assist in selecting, extending, or designing {standard} concrete implementations that, although
potentially different in detail, will nonetheless capture the same necessary features of a
smart legal agreement. Different standard concrete formats might for example differ in choices
such as character encoding and hashing format.  Standard concrete formats could also
facilitate, for example, automatic analyses across a wide range of smart legal
agreements.

\item
{\em Ontologies.\footnote{See
    \url{https://en.wikipedia.org/wiki/Ontology_(information_science)}}} Standard formats
  include not only standard syntax but also the use of standard ontologies.  Existing standards
  such as the Enterprise Data Management Council's Financial Industry Business Ontology (FIBO)
  \cite{fiboedm} could be leveraged to assist semantic analysis of legal prose.\footnote{FIBO
    provides a vocabularly of terms using two forms of definition \cite{fibo-foundations} for
    each concept: (i) a structured ontology specification of the concept, and its relationships
    to others, represented using the Web Ontology Language (OWL), and (ii) a natural language
    definition which represents the concept using the vocabulary of the finance industry.}  As
  noted in \cite{fibo-foundations}, FIBO can be utilised to perform semantic reasoning and aid
  the development of querying applications.

The FIBO specifications define, among other things, legal and business entities, instruments,
products, services, interest rates, currency exchange rates, economic indicators and market
indices.  Textual markup (discussed later) could be extended to support semantic analysis and
reasoning using OWL, but the details of such extensions are beyond the scope of this paper.

\item
{\em Binding smart legal agreement and code.} This comprises two aspects:
\begin{enumerate}
\item passing execution parameters to the smart contract code to direct its operation;
\item providing a succinct way to identify the legal agreement uniquely at an operational level
  to support finding the legal agreement if needed for review or dispute resolution.
\end{enumerate}

A candidate solution for the requirements of an operational-level unique agreement identifier
is a cryptographic hash of the smart legal agreement that is passed as a parameter to the smart
contract code and stored in the instantiated smart contract; this is the technique used in
Ricardian Contracts. Note there are other similar solutions, such as Monax Industries' ``dual
integration'' \cite{monax} which additionally provides a reverse link by adding a unique
identifier for the instantiated smart contract to the final smart legal agreement.
\end{enumerate}

\section{Abstract core specification}
\label{sec:minimal-abstract-specification}
The above essential requirements support the two legs of our previous definition of a smart
contract --- i.e. that it is both automatable and enforceable, where enforcement occurs via
legal enforcement\footnote{ Further discussion on the legal enforceability of smart contracts
  can, for example, be found in \cite{nortonrose}.} of rights and obligations
\cite{SCT2016}.

In this section, we start to explore the design landscape of a potential serialised format for
storage, retrieval and transmission of smart legal agreements. We present an \linebreak {\em abstract
  specification} that defines the logical structure of smart legal agreements --- and divide
our presentation into two parts:
\begin{enumerate}
\item
a small {\em core} specification (in this section), which is sufficiently general that it can
serve as the basis for a wide range of possible specifications, and
\item
a longer discussion (in the next section) of possible {\em design options}, with illustrative
example specifications which are not intended to be prescriptive --- allowing the final choices
on these matters to be made later (e.g. by standards bodies).
\end{enumerate}

\subsection{Notation}
We use the BNF-like\footnote{See \url{https://en.wikipedia.org/wiki/Backus-Naur_form}} notation
summarised in Figure~\ref{fig:bnf} below, with the exceptional semantics that the elements are
unordered.  For example, ``a ::= b c'' defines ``a'' as ``b c'' or ``c b'', and ``x ::= y* z*''
defines ``x'' as any combination of zero or more of the elements ``y'' and ``z''.

\begin{figure}[h]
\begin{center}

{\begin{tabular}{c l }
\hline 
\ \ &\ \ \\
::=&Is defined as \\
$|$&Or\\
$*$&Zero or more occurrences\\
$+$&One or more occurrences\\
\ \ &\ \ \\
\hline
\end{tabular}}

\parbox{4.9in}{\caption{\footnotesize{Notational conventions used in this
      paper.}}\label{fig:bnf}}
\end{center}
\end{figure}

\vspace{-5mm}
\subsection{Representation of smart contracts}
\label{representation-of-the-ricardian-contract}

Inspired by the Ricardian Contract triple of ``prose, parameters, and code''
\cite{grigg2004ricardian, grigg2015sumofchains}, we define a {\em core} abstract specification
that represents: (i) interim drafts of a contract (including the empty starting state), (ii)
the final version of a contract, and (iii) a smart contract comprising multiple smart legal
agreements and/or smart contract code implementations. Many detailed specifications can be
derived from the following core specification\footnote{For example, if only an electronic
  confirmation document is available then we might have no {\em legal-prose} element, one
  {\em parameters} element and no {\em agreement-header} element.}.

\begin{equation}\tag{D1}\label{eqn:smart-contract}
  \operatorname{smart-contract} ::= \operatorname{smart-legal-agreement* \; \; smart-contract-code*}
\end{equation}

\begin{equation}\tag{D2}\label{eqn:smart-legal-agreement}
  \operatorname{smart-legal-agreement} ::= \operatorname{legal-prose* \; \;parameters* \;
    \;agreement-header*}
\end{equation}

\vspace{2mm}

\section{Options in the design landscape}
\label{sec:options-in-the-design-landscape}
Beyond the ``core'' abstract specification given above, which we believe to be generic, any
further abstract specification opens up a landscape of possible design options.  The
predominant activity in this landscape is the identification and recording of metadata in a way
that best fits the requirements of smart contracts.  We have previously stated that the
specific choices on these matters should be made later; however, we will attempt to describe
many of the design options that arise, and to provide illustrative examples of how particular
choices might be captured in a specification.

The illustrative examples given below do not constitute preferences or suggestions; each is
provided merely to clarify one or more aspects of the discussion.

\subsection{Markup}
In our abstract specification, we wish to encode the logical structure of smart legal
agreements. Without intending to be prescriptive in this matter, we focus on textual
representation. Furthermore, in this paper we focus on the use of static textual markup rather
than data transformations and procedures that are also important for workflow processes.

Markup can be either attached to text or associated with a position in text (for example an
``anchor'' used for cross-referencing). There are many different forms of textual markup, for
example:
\begin{itemize}
  \item
Descriptive markup can, among other things, refer to the structure of the text (such as
``heading'' or ``section'' or a position in the text) and/or the meaning of the text (such as
``parameter'' or ``indemnity clause''). 

\item
Presentational markup refers to how the marked-up text should be rendered (such as ``bold'' or
``italic''). This could be implemented as: (i) inline presentational markup
attached to text within legal prose, or (ii) style sheets that map descriptive
markup to presentational markup.  Formatting is a key aspect of legal documents, and is
therefore an important part of an abstract specification.
\end{itemize}

\noindent
We can specify the above example of two forms of markup as follows:

\begin{equation*}\label{eqn:markup}
  \operatorname{markup \; ::= \; presentational-markup \; | \; descriptive-markup}
\end{equation*}

\vspace{1.5mm}
\noindent
In the following sections, we explore the design options for inline textual markup of smart
legal agreements.

\subsection{Design options for prose}
\label{sec:design-choices-for-prose}

\subsubsection{Text}
\label{sec:text}
There are different ways to view legal prose.  For example, it may be viewed as a linear
sequence of pieces of text, each of which is either marked-up or is not, and where there may be
markup occurring between pieces of text.  Another way is to view legal prose as a hierarchical
structure of elements, such as one or more parts (recitals, definitions, schedules, etc) each
of which contains one or more paragraphs, themselves containing sentences that contain words,
and so on.  In the latter example, markup could be applied to each of the hierarchical elements
and there may be markup immediately preceding or immediately following an element.

Many different abstract specifications of {\em legal-prose} are possible, for example:

\begin{equation*}\label{eqn:prose}
  \operatorname{legal-prose \; ::= \; \; text* \;\; markup* \;\; text-with-markup*\;}
\end{equation*}

\vspace{1.5mm}
\noindent
This motivates the choice of a specification for {\em text-with-markup}.  We give one example
below, to illustrate how this could be achieved, but there are many other ways:

\begin{equation*}\label{eqn:text-with-markup}
  \operatorname{text-with-markup \; ::= \; markup^+ \; text \;}
\end{equation*}

\noindent

\subsubsection{Lists and tables}

Legal text often uses various layout devices such as lists and tables.  Lists have both a
presentational aspect (e.g. they may be numbered, bulleted, or dashed) and a logical aspect
(e.g. it is possible to refer to list items by position when a cross-reference is made from
elsewhere in the text).  Tables may also be referred to by number, and may have additional
caption text.

In order to detect certain kinds of error syntactically, lists and tables may require special
markup rules.  For example, if a list item were to appear outside a list this would normally be
detected as a syntactic error.  Although other kinds of markup such as bold and italic can be
nested, normally there would be no rules that permit one to be nested inside the other but not
vice-versa.  If a design requirement is to detect list and table errors syntactically then, in
the abstract specification, lists will be treated in a different way to ``simple'' markup (and
for the same reason tables may also be treated differently).  There are many ways to achieve
this in a specification: one way might for example require {\em text-with-markup} to be given a
more complex definition, with a layered structure to capture the legal nesting of list items
inside a list.  An elegant method might depend on the notation being used (e.g. whether
the notation permits a recursive definition).  

\subsubsection{Cross-references} 
\label{sec:references}
Cross-references are a common feature in legal text, where a reference to a target is embedded
inside a source piece of text.  The target may be either a referenceable piece of text (such as
an item in a list, a table number, or some text with special markup), or a referenceable
position in the text --- where ``referenceable'' means in each case that it must have some kind
of identifier.  The former are sometimes known as ``segments'' and the latter as
``anchors''.\footnote{See \url{http://www.tei-c.org/Vault/P4/Lite/U5-ptrs.html}}

When editing or viewing a large document, it is useful to be able to jump from the source of a
cross-reference to its target and then to jump back again (see also
Section~\ref{sec:inter-document-cross-references}).  It is also very useful to know whether a
piece of text is the target of one or more cross-references (especially if the target is to be
edited).  For these reasons, there might be a design requirement for cross-references to be
bidirectional.\footnote{There are other useful requirements that might additionally be applied
  to cross-references and should be addressed as design options.  For example: each
  cross-reference should be unique within the agreement, each should have a single source and a
  single target, jumping between source and target (or vice versa) should be fast, a section of
  text may be the source for many references and may be the target for many references, sources
  and targets may be nested but may not be overlapped, etc.}  There are many ways to specify
bidirectional cross-references.  Two examples are:

\begin{enumerate}
  \item as an inline markup applied to both the source and the target of the
    cross-reference, with full information about the source and the target being attached to
    both the source and the target;
  \item as a small inline markup applied to both the source text and the target text,
    each holding a unique identifier for the cross-reference; with a list or table of all
    cross-references and full details of their sources and targets being held in the {\em
      agreement-header}.
\end{enumerate}

\noindent
Both of the examples given above would require specific descriptive markup to be used both for
the source and for the two different types of target.

\subsubsection{Redacted text}
Redacted text, for example proprietary or privileged text, should not be printed out or
transmitted to a third party --- such text is typically formatted in a blacked-out or
whited-out fashion in a redacted copy.  This may be specified in a variety of ways, for example
there may be specific markup to indicate that text is either ``To Be Redacted'' (and perhaps
also ``Has Been Redacted'').

\subsubsection{Optional clauses}

During the drafting of legal text, the author may wish to search for standard clauses to
insert, or there may be a requirement to provide the author with a template that includes
embedded optional clauses.  The former requirement falls mostly outside the scope of specifying
a standard format for storage or transmission of a smart legal agreement (since a collection of
standard clauses need not be a ``smart legal agreement'').  The latter requirement would
require an extension to the specification, to include a ``choice'' element that would contain
several pieces of text from which the author should choose.  The design options are similar to
those discussed for lists and tables; the requirement could be achieved in many ways, one of
those being to treat the various choices as being similar to a list (with special syntactic
rules like a list) but with a different variant of descriptive markup.

\subsection{Design choices for parameters}
\label{sec:design-choices-for-parameters}

An abstract specification for smart legal agreements must cater for many different scenarios,
especially with regard to parameters. These parameters might initially only be embedded in the
{\em legal-prose} element of the smart legal agreement
(Definition~\ref{eqn:smart-legal-agreement}) --- for example, if only the legal documentation
is available at the start. Alternatively, the parameters might initially only be included in
the {\em parameters} element --- for example, if a confirmation document is available, but the
associated legal documentation has not yet been included.

Later in the lifecycle of the smart legal agreement, several design options are available
including:

\begin{itemize}
\item that all parameters should be identified within the {\em legal-prose} using inline markup
  and that the {\em parameters} element should not be used;
\item that all parameters should be held in the {\em parameters} element and that wherever the
  prose contains text describing parameters they should not be marked up as parameters;
\item that all parameters should be identified within the prose using inline markup {\bf and}
  information about each parameter should also be kept in the {\em parameters} element
  (e.g. for operational convenience).
\end{itemize}

\noindent
Parameters that are embedded in the prose may not initially be identified and so it is
important that we are able to identify and retrieve parameters from the legal prose.  The key
data that we need for each parameter is its value.  Parameters are also typically referenced by
name, which should also be recorded.

\subsubsection{Parameter data types}

A concrete implementation could choose between (i) a mono-typed system for parameters where
every parameter has the same type (e.g. text), and (ii) a typed system where the
identification of a parameter entails identification of its {\bf name}, {\bf value} and {\bf
  type} (where the available types would be determined by the concrete implementation).

If a smart legal agreement is transmitted to a counterparty, it may be necessary to include in
the serialised format a list of all non-standard type definitions.  This could be held in the
{\em agreement-header} element.

Complex parameter types such as arrays, lists and expressions could also be supported.  One
design decision could be to utilise a compositional description language to create business
logic expressions that could be identified as parameters.

\subsubsection{Identification of parameters}

There are several different possible scenarios for the development of smart legal agreements.
For example, as discussed above there is a design choice as to whether
parameters are or are not held in the {\em parameters} element.  It is possible that an
execution parameter may be held in the {\em parameters} element without appearing in the prose
(there may not be a {\em legal-prose} element).  It is also possible that the execution
parameters and their values might appear in the prose and not be held in the {\em parameters}
element.  Finally, it is possible that execution parameters might appear in both the {\em
  legal-prose} and the {\em parameters} elements.

If parameters appear in the prose, it is essential that there be a way to identify those
parameters.  They must also be retrievable so that they can be passed to the code when it
executes.  A design choice exists in how to identify the parameters. This
could be achieved in many ways but we give two examples below:

\begin{enumerate}
\item
One design decision might be to attach markup to each piece of text in {\em legal-prose} that
provides parameter information (the markup would for example capture the name, type and value
of the parameter in each case), yet leave the {\em parameters} element empty.  When it is time
to execute the code the prose can be searched to find the names, types and values of all
execution parameters and these can be communicated to the code.
\item
Another design decision might be to attach markup to each piece of text that provides parameter
information, but store the name, type and value information for each parameter in the {\em
  parameters} element.  The markup in the prose could, if required, store an identifier
referencing the parameter data in {\em parameters}.  When it is time to execute the code, all
execution parameters in the {\em parameters} element can be communicated to the code without
needing to search the prose.  The {\em parameters} element may be useful when using
standardised methods that require parameters to be presented separately from the prose, in
which case a further design decision might be to make the {\em parameters} isomorphic to a
standard format (e.g. by using FpML \cite{fpml} as a concrete implementation).
\end{enumerate}

\noindent
The identification of parameters in the prose, when deemed necessary, could be specified as
descriptive markup. There are also many ways that parameter data could be held in the {\em
  parameters} element. An example specification is shown below:

\begin{equation*}
  \operatorname{parameters\; ::= \; parameter*}
\end{equation*}
\begin{equation*}\label{eqn:parameter-data}
  \operatorname{parameter\;\; ::= \;\; parameter-name \;\;parameter-type \;\;parameter-value}
\end{equation*}

\subsubsection{Other kinds of parameter} 

Legal prose may contain important data that is used for purposes other than execution of the
smart contract code.  For example, data may be used for compliance reporting or may be part of
a definition that is important in a legal sense but is not needed by the smart contract code.

One design choice would be to let such data be identified each time it is
used, via a search of the legal prose.  However, it might be deemed be advantageous to identify
this data and keep a separate record of it for ease of reference (for example, by analysis and
reporting systems).  This record could potentially be held in the {\em agreement-header}
element.

\subsection{Cryptographic hashing}
\label{sec:hash}
A cryptographic hash\footnote{See
  \url{https://en.wikipedia.org/wiki/Cryptographic_hash_function}} is the output of a one-way
  mapping from data of arbitrary size to a bit string of fixed (and typically small) size.  It
  is ``one-way'' because it is not feasible to obtain the original data from the hash.  The
  same data always gives the same hash, and a small change in the data can lead to a large
  change in the hash.  Furthermore, in general it is not feasible to find two pieces of data
  with the same hash.

Hashes could be used in various ways.  For example: (i) as a unique identifier for a smart
legal agreement --- as the value of an execution parameter passed to the smart contract code
and/or as an index into a repository, (ii) as a method to detect modification of a smart legal
agreement after it has been signed, or (iii) as a method to detect modification of a
pre-authorised piece of text (e.g. a legal clause) that has been used inside a smart legal
agreement. These techniques can also be used to evidence data tampering.

The Ricardian Contract uses a cryptographic hash of the entire document as an identifier.  In
general terms, using a cryptographic hash requires a canonical form of the document to avoid
generating ``false positive'' modification alerts from semantically equivalent forms (such as
alternative nesting of markup, e.g. {\em Bold Italic} versus {\em Italic Bold}).\footnote{See
  for example the section on ``XML canonicalisation'' at
  \url{https://en.wikipedia.org/wiki/XML_Signature}}

Whether to use hashes for these or other purposes is a design option, as is the decison of
where to store the hash.

\subsection{Structure, header and linking to code}
\label{sec:design-options-for-document-structure}

There are several design options that relate to the overall structure of a smart legal
agreement, to its metadata, and to the linking of prose to smart contract code.  We discuss
these below.

\subsubsection{Separating parts of the agreement} 

Large agreements may benefit from separation into logically separate parts (e.g. definitions,
schedules and annexes).  This can be achieved in many ways:

\begin{itemize}
\item
through use of {\em markup} in the prose to identify the start and end of each new part;
\item
by defining {\em legal-prose} in a hierarchical way as discussed in Section~\ref{sec:text};
\item
by representing a {\em smart-legal-agreement} as having multiple {\em legal-prose} elements
(possibly with multiple {\em agreement-header} elements);
\item
by representing a {\em smart-contract} as having multiple {\em smart-legal-agreement} elements.
\end{itemize}

%% \noindent
%% These design options are closely related to the consideration of
%% multi-document agreements, which we will discuss in Section~\ref{sec:multi-document-agreements}.

\subsubsection{Document header}
\label{sec:design-options-for-document-administration}

We have previously identified several design options where it might be advantageous to hold
information in the {\em agreement-header}.  A wide range of information could be held;
generally, this would be information that either does not exist in the prose, or for which it
is administratively easier if a copy of that information is also held in a header.  Examples
might inlude:

\begin{itemize}
\item A list or table of all cross-references and full details of their sources and targets.
\item A list of all non-standard type definitions.
\item Various dates (dates of signing, execution date, effective date, and so on).
\item Digital signatures.
\item A cryptographic hash of the smart legal agreement (see Section~\ref{sec:hash}).
\item Various identifiers for the smart legal agreement, such as a local filing identifier. A
  cryptographic hash could also be used as a globally-unique identifier (and potentially also
  usable for local filing if desired).  The issue of identification is wide-ranging --- for
  example, an agreement may have a globally-unique identifier and an individual trade may have
  a mandated trade identifier.
\item A style sheet for presentational formatting.
\item An edit history and version control data --- this is discussed further in
  Section~\ref{sec:edithistory}.
\end{itemize}

\noindent
Although metadata is essential throughout the lifecycle of a smart contract, it may be
necessary to remove certain metadata in a final version of the contract.

\subsubsection{Binding with code}
\label{sec:design-options-for-linking-to-code}

There are several design choices to be considered regarding binding the legal prose with the
smart contract code:
\begin{itemize}
\item There may be several different instances of standardised code that could run the smart
  contract (for example, corresponding to different code versions or different possible
  execution platforms). Note that the specification of {\em smart-contract} (in Definition
  \ref{eqn:smart-contract}) permits multiple instances of {\em smart-contract-code}.

\item When smart contract code is instantiated onto an execution platform, there should be a
  mechanism for passing the execution parameters to that code. In addition, there should be a
  method for passing a unique identifier for the smart legal agreement to that code and the
  execution platform should embody a method to make that identifier available.

\item After the smart contract code has been instantiated on an execution platform, there may
  be a requirement to store within the smart legal agreement a unique identifier of that
  executing instance of the code (e.g. see ``dual integration'' described at the end of
  Section~\ref{sec:essential-requirements}). This unique identifier could, for example, be
  stored in the {\em agreement-header}, {\em parameters}, or {\em legal-prose}. According to
  the governance procedure, this may constitute a change to the agreement that might require a
  further level of authorisation before proceeding with execution of the smart contract code.
\end{itemize}

\section{Design options for multi-document agreements}
\label{sec:multi-document-agreements}

In this section we consider further design options that relate specifically to agreements that
comprise multiple documents.  We also consider what options may arise where there is a
hierarchical relationship between, for example, standard templates, local templates, agreements
and trades.  Finally, we consider workflow topics such as an edit history and versioning.

\subsection{Document groups}
There are many ways in which smart contracts comprising multiple documents could be
specified. The simplest specification is for each document to be a separate {\em
  smart-legal-agreement}. An alternative would be for each document to be a separate {\em
  legal-prose} within a {\em smart-legal-agreement} potentially containing shared {\em
  parameters} and/or {\em agreement-header}.

\subsection{Document types and status}
Where an agreement comprises multiple documents, typically each document has a well-defined
role or ``type''.  For example, with agreements for financial derivatives there may be a Master
Agreement, Schedule, Credit Support Annex, and so on. The document type could, for example, be
held in an appropriate {\em agreement-header} element.

Furthermore, it might be desired to keep track of the status of each document (for example,
whether parameters have been identified, whether the legal prose has been agreed with
counterparties, and so on).  There may be many ways to do this; one design option would be to
store a ``document status'' inside an appropriate {\em agreement-header} element.

\subsection{Document hierarchies}
In \cite{SCT2016} we proposed that the lifecycle of a smart legal agreement would start with a
Smart Contract Template.  Organisations may also wish to develop local versions of the
templates produced by standards bodies.  Furthermore, there might be a tree hierarchy of local
versions developed for different purposes.  This hierarchy is then conceptually expanded
downwards as agreements are derived from templates, and as trades are derived from
agreements.\footnote{Additionally, there might be a design requirement for a hierarchical
  precedence of documents.}

As a design option, the notion of a document ``type'' could be used to record for example
whether a document is an industry-standard template, a local template, an agreement, and so on.
However, as a further design option there may be a desire to record at what level a given
document exists in the document hierarchy.  This might be specified in many ways; one example
would be to extend the {\em agreement-header} element with (i) a set of identifiers for parent
documents (one level higher in the hierarchy, from which this document has been derived), and
(ii) a set of identifiers for child documents (one level lower in the hierarchy).
Industry-standard templates could exist at the top level of this hierarchy and have no parents.

\subsection{Inter-document cross-references}
\label{sec:inter-document-cross-references}
Cross-references from a piece of text in one document to a different target document are common
in legal prose; this may be a reference to a piece of text, to a locaton within the text, to
the target document itself, or even to an entire agreement.  If there is a requirement to
support these cross-references so they can be quickly navigated from source to target and back
again, then these inter-document cross-references should also be bidirectional.

We have previously discussed the design options for cross-references in
Section~\ref{sec:references}.  Additional design options could include cross-references between
different documents (which may or may not be in the same smart legal agreement) and different
kinds of document identifier (including local and globally-unique identifiers).

Furthermore, where a document is the target of many inter-document cross-references, it could
be required that any editing of the target document that would cause target pieces of text to
move should not cause any changes in documents that contain the sources of those
cross-references. This might be a particularly important design option if the source has been
previously negotiated, agreed and hashed.\footnote{An example specification to achieve this
  might maintain outgoing and incoming indirection tables in the {\em agreement-metatdata}
  element for each document --- the source would then refer to a fixed entry in the metadata of
  the target regardless of any movement of the target within its document.}

\subsubsection{Incremental parameter definitions}
Incremental parameter definition refers to the common practice of declaring a name in one
document and then giving that name a value (``binding'' the name to a value) in a different
document.  Furthermore the name, and by extension its value, may be referenced by a third
document.\footnote{The topic of parameter scope merits further discussion,
  e.g. the visibility of parameter names (and their values) outside a smart legal agreement.}
For example, a parameter name might be declared in a Master Agreement, given a value in a
Schedule, and then used in a Confirmation.

As a design option, this feature of incremental definition of parameters could be applicable
within all types of document including for example templates as issued by standards bodies,
locally-modified templates, agreements, and so on.  Any document could then be considered to be
``parameterised'', to the extent that it uses names whose values are to be provided in another
document and/or at a later time.

Incremental parameter definitions could be specified in different ways.  One example would be
to start by permitting {\em parameter-value} to have a special value such as {\em unbound}
(meaning that although the name has been declared there is as yet no value).\footnote{This
  might be useful when defining templates, so that a value is initially {\em unbound} but
  is then changed to be a specific value when for example the template is used to create an
  agreement.}  A second special value such as {\em binding-location} might be set at a later
time to indicate that a value has been provided in another document (perhaps with an inter-document
cross-reference to that value).

\subsubsection{Version control and edit history}
\label{sec:edithistory}
Version control ranges from the simple recording of a version number and timestamp for each
document (and for the agreement as a whole), to the complex tracking of versions for multiple
documents with multiple branches (for example different branches might be created for
locally-stored and transmitted versions, so that sensitive metadata is not transmitted).  By
contrast, an edit history maintains a complete log of all changes to a document (e.g. for audit
purposes).  For smart legal agreements the key design options include (i) recording the current
version number and timestamp, (ii) keeping a complete log of changes --- possibly including rejected
amendments, approvals, and counterparty communication, and (iii) designing a branching and
merging strategy for versions, so that exported (transmitted) versions can be re-imported.
      
The version number and date for documents can be stored in an appropriate {\em
  agreement-header} element, as can the document edit history.

\section{Summary and further work}

\subsection{Summary}
This paper began by presenting what we believe are the essential requirements for smart legal
agreements, covering creation and editing, standard formats, legal execution protocols, binding
to smart contract code, and making them available in acceptable forms. We then provided an
abstract core specification for a smart contract and a smart legal agreement.

We then explored the design landscape for a seralised format for storage and transmission of
smart legal agreements, including markup for metadata, design options for prose (lists, tables,
cross-references, redacted text and optional clauses), design choices for parameters (data
types and identification), cryptographic hashing, and multi-document agreements (groups, types,
hierarchies, and incremental parameter definitions).

Our aim is to support the financial services industry in exploring how legal prose can be
connected with parameters and code, and trade associations when reviewing existing data
standards to take account of the features of smart legal agreements.  This work therefore
provides a preliminary step towards supporting industry adoption of legally-enforceable smart
contracts.

\subsection{Further work}

There are many design options that we have not yet investigated, for example: 
\begin{itemize}
\item
permissioning and access control, ranging from the entire document to specific document portions
such as clauses and sentences;
\item
roles and responsibilities of those able to operate on an agreement (e.g. designated signatories);
\item
the ability to specify that if a given clause is included, or modifed, then the agreement must
be referred for specific authorisation;
\item
a more detailed exploration of whether/how existing standards (such as FpML) might be extended
beyond key-value pairs towards support for the negotation of smart legal agreements including
legal prose;
\item
  financial transactions that are cleared via a central counterparty;
\item
  the treatment of discretionary rights within an agreement.
\end{itemize}

\noindent
It is also important to consider the workflow and system requirements for Smart Contract
Templates. Any concrete implementation of a ``standard'' format will be used for transmission
and storage between and within systems that will support substantial workflow (including
negotiation between the parties to each agreement). There are some foundational architecture
topics that arise for any such system, including many previously raised in
\cite{palmersctarchitecture} such as issues surrounding template and agreement governance,
repositories, and jurisdiction.

There remain many questions and design decisions to be explored.  This will require substantial
work and collaboration by the financial services industry, standards bodies,
academia\footnote{For example, further development of the CLACK language \cite{SCT2016} will be
  pursued at University College London.}, and lawyers.

% Hyphens back to normal
\pretolerance=-1
\tolerance=-1
\emergencystretch=0pt

\bibliography{SCTERDO2016}

\begin{thebibliography}{10}

\bibitem{axoni}
Axoni.
\newblock {Axoni Core}, 2016.
\newblock \url{https://axoni.com/}.

\bibitem{smartcontracttemplates}
L.~Braine.
\newblock {Barclays' Smart Contract Templates}, 2016.
\newblock Barclays London Accelerator, \url{https://vimeo.com/168844103/} and
  \url{http://www.ibtimes.co.uk/barclays-smart-contract-templates-heralds-first-ever-public-demo-r3s-corda-platform-1555329/}.

\bibitem{SCT2016}
C.D Clack, V.A Bakshi, and L.~Braine.
\newblock {Smart Contract Templates: foundations, design \linebreak landscape
  and research directions}.
\newblock {\em The Computing Research Repository (CoRR)}, abs/1608.00771, 2016.
\newblock Also at arXiv.org: \url{http://arxiv.org/abs/1608.00771/}.

\bibitem{clausematch}
{ClauseMatch}, 2016.
\newblock \url{http://www.clausematch.com/}.

\bibitem{commonform}
{Common Form}, 2016.
\newblock \url{https://commonform.org/}.

\bibitem{commonaccord}
{CommonAccord}, 2016.
\newblock \url{http://www.commonaccord.org/}.

\bibitem{corda}
Corda, 2016.
\newblock \url{https://www.corda.net/}.

\bibitem{digitalasset}
{Digital Asset}.
\newblock {The Digital Asset Platform - Non-technical White Paper}, 2016.
\newblock Available at
  \url{https://digitalasset.com/press/digital-asset-releases-non-technical-white-paper.html}.

\bibitem{json}
{ECMA International}.
\newblock {The JSON Data Interchange Format}, 2016.
\newblock
  \url{http://www.ecma-international.org/publications/files/ECMA-ST/ECMA-404.pdf}.

\bibitem{fiboedm}
{Enterprise Data Management Council (EDM Council)}.
\newblock {Financial Industry Business \linebreak Ontology}, 2016.
\newblock \url{http://www.edmcouncil.org/financialbusiness/}.

\bibitem{ethereum}
Ethereum, 2016.
\newblock \url{https://www.ethereum.org/}.

\bibitem{fpml}
{Financial products Markup Language (FpML)}, 2016.
\newblock \url{http://www.fpml.org/}.

\bibitem{grigg2004ricardian}
I.~Grigg.
\newblock {The Ricardian Contract}.
\newblock In {\em Proceedings of the First IEEE International Workshop on
  Electronic Contracting}, pages 25--31. IEEE, 2004.
\newblock \url{http://iang.org/papers/ricardian_contract.html}.

\bibitem{grigg2015sumofchains}
I.~Grigg.
\newblock {The Sum of All Chains --- Let's Converge!}, 2015.
\newblock Presentation for Coinscrum and Proof of Work.
  \url{http://financialcryptography.com/mt/archives/001556.html}.

\bibitem{hotdocs}
{HotDocs}, 2016.
\newblock \url{http://www.hotdocs.com/}.

\bibitem{hyperledger}
{Hyperledger Fabric}, 2016.
\newblock \url{https://github.com/hyperledger/fabric/}.

\bibitem{monax}
Monax Industries.
\newblock {Dual Integration}, 2016.
\newblock \url{https://monax.io/explainers/dual_integration/}.

\bibitem{openxml}
ECMA International.
\newblock {Standard ECMA-376 - Office Open XML File Formats}, 2016.
\newblock
  \url{http://www.ecma-international.org/publications/standards/Ecma-376.htm}.

\bibitem{fibo-foundations}
{Object Management Group (OMG)}.
\newblock {Financial Industry Business Ontology - Foundations \linebreak
  (FND)}, 2016.
\newblock OMG Document Number: dtc/2016-03-03,
  \url{http://www.omg.org/spec/EDMC-FIBO/FND/1.1/Beta2/PDF/index.htm}.

\bibitem{opendocument}
{Organization for the Advancement of Structured Information Standards (OASIS)}.
\newblock {Open Document Format for Office Applications (OpenDocument) Version
  1.2}, 2011.
\newblock \\
  \url{http://docs.oasis-open.org/office/v1.2/OpenDocument-v1.2.pdf}.

\bibitem{palmersctarchitecture}
N.~Palmer.
\newblock {Architectural Considerations for Smart Contract Templates}.
\newblock In {\em The R3 Smart Contract Templates Summit}, pages 90--92, June
  2016.
\newblock
  \url{http://r3cev.com/s/R3-Smart-Contract-Templates-Summit-_FINAL.pdf}.

\bibitem{nortonrose}
R3 and {Norton Rose Fulbright}.
\newblock {Can smart contracts be legally binding contracts?}, 2016.
\newblock
  \url{http://www.nortonrosefulbright.com/knowledge/publications/144559/can-smart-contracts-be-legally-binding-contracts/}.

\bibitem{smartdx}
{Smart Communications}.
\newblock {SmartDx}, 2016.
\newblock \url{https://www.smartcommunications.com/products/smart-dx/}.

\bibitem{stark2016}
J.~Stark.
\newblock Making sense of blockchain smart contracts, 2016.
\newblock \url{http://www.coindesk.com/making-sense-smart-contracts/}.

\bibitem{contractexpress}
{Thomson Reuters}.
\newblock {Contract Express Author}, 2016.
\newblock \url{http://www.contractexpress.com/}.

\bibitem{xml}
{W3}.
\newblock {Extensible Markup Language (XML)}, 2008.
\newblock \url{https://www.w3.org/TR/xml/}.

\end{thebibliography}
\bibliographystyle{plain}

\end{document}